\documentclass[12pt, a4paper]{article}
\pdfoutput=1
\usepackage{amsmath}
\usepackage{amssymb}
\usepackage{amsfonts}
\usepackage{slashed}
\usepackage{graphicx}
\usepackage{color}

\begin{document}

\newcommand\lsim{\mathrel{\rlap{\lower4pt\hbox{\hskip1pt$\sim$}}
    \raise1pt\hbox{$<$}}}
\newcommand\gsim{\mathrel{\rlap{\lower4pt\hbox{\hskip1pt$\sim$}}
    \raise1pt\hbox{$>$}}}

\definecolor{Red}{rgb}{1.,0.,0.}
\newcommand{\Red}[1]{{\color{Red}{#1}}}

\long\def\symbolfootnote[#1]#2{\begingroup%
\def\thefootnote{\fnsymbol{footnote}}\footnote[#1]{#2}\endgroup}

\allowdisplaybreaks[1]

\thispagestyle{empty}

\sf
\centerline{\Huge Lepton flavor violation in type I + III seesaw}

\vspace{7mm}

\centerline{\large Jernej F. Kamenik$^{1,3}$ \symbolfootnote[1]{e-mail: jernej.kamenik@lnf.infn.it} and Miha Nemev\v sek$^{2,3}$ 
\symbolfootnote[2]{e-mail: miha.nemevsek@desy.de}}

\vspace{1mm}

\centerline{ {\it\small $^1$ INFN, Laboratori Nazionali di Frascati, I-00044 Frascati, Italy} }
\centerline{ {\it\small $^2$ II. \ Institut f\"ur Theoretische Physik, Universit\"at Hamburg,} }
\centerline{ {\it\small
Luruper Chaussee 149, 22761 Hamburg, Germany} }
\centerline{ {\it\small $^3$ J.\ Stefan Institute, 1000 Ljubljana, Slovenia} }
\vspace{5mm}

\centerline{\large\sc Abstract}
\begin{quote}
\small
In the presence of a low scale seesaw of type I + III, flavor violating effects in the leptonic sector are expected. Their presence in the charged sector is due to the mixing of the fermionic vector-like weak triplets with the chiral doublets, which cause non-universality of the tree-level $Z$ coupling. We investigate the bounds on the Yukawa couplings which are responsible for the mixing and present the results for two minimal cases, a fermionic triplet with a singlet or two fermionic triplets. Different channels for these processes are considered and their current and future potential to probe these couplings is discussed. 
\end{quote}
\rm

\section{Introduction}
\label{secIntro}

Available experimental data on neutrino oscillations indicates a small mass of left-handed neutrinos. This is in contrast with the Standard Model (SM) where the left-handed neutrinos are massless. Also, the nature of neutrinos, whether they are Dirac or Majorana particles, is not known. The latter possibility is theoretically most compelling, since it introduces new physics at the scale $\Lambda$, where the neutrino mass operator
\begin{equation} \label{eqWeinberg}
  \mathcal O_\nu^{d=5} = y_\nu^{ij} \, \frac{L_i H L_j H}{\Lambda}
\end{equation}
is formed. There are only three different ways to realize this operator at the tree level when a single representation is added \cite{Ma:1998dn}.

Adding a right-handed neutrino is referred to as the type I seesaw \cite{Minkowski:1977sc, Yanagida:1979as, GellMann:1980vs, Glashow:1979nm, Mohapatra:1979ia}, while an extra bosonic triplet with hypercharge 1 results in type II seesaw \cite{Magg:1980ut, Lazarides:1980nt, Mohapatra:1980yp}. The third option is to couple the leptonic and Higgs doublets to a fermionic weak triplet with zero hypercharge and this is the type III seesaw \cite{Foot:1988aq}. 

Unfortunately, the scale $\Lambda$ is not known since it depends on the size of the Yukawa couplings. If they are of order one, as in certain Grand Unified Theories (GUTs), Eq.\eqref{eqWeinberg} predicts $\Lambda$ around $10^{13}$ GeV. Such a high scale would make it very hard to probe the origin of the mass operator directly at a collider. On the other hand, when Yukawa couplings in Eq.\eqref{eqWeinberg} are small, the seesaw scale may lie anywhere below $10^{13} \text{ GeV}$. Notice also, that the small Yukawa couplings are technically natural due to a protective chiral symmetry.

Recently, a grand unified model has been proposed \cite{Bajc:2006ia} which is an extension of the minimal Georgi-Glashow $SU(5)$ model with a fermionic adjoint representation, that predicts a low mass for a fermionic triplet from unification requirements \cite{Bajc:2007zf}. Neutrino masses are realized with a combination of type I and III seesaws together with an upper bound on the fermionic triplet around TeV. Due to the fact that they are coupled to electroweak gauge bosons, one can produce the fermionic triplets at the LHC and measure the origin of neutrino masses by studying their decays \cite{Bajc:2007zf, Franceschini:2008pz, Arhrib:2009mz}. In principle, one can even distinguish various seesaw types at a collider by studying events by their charged lepton multiplicity \cite{delAguila:2008cj}.

Besides the neutrino mass operator, also higher dimensional operators are produced below the seesaw scale, the size of which can be constrained by flavor changing processes \cite{Abada:2007ux}. While the leptonic mixing matrix becomes non-unitary in both type I and III cases \cite{Antusch:2006vwa}, the unique feature of the type III is the presence of charged lepton flavor changing neutral currents (FCNCs) at the tree level. Our aim is to establish, whether testing such processes may shed some light on the origin of neutrino mass in a minimal model with a predicted light triplet below TeV.

In contrast to previous studies \cite{He:2009tf, Arhrib:2009xf}, we use the existing data from neutrino oscillation experiments to express Yukawa couplings which enter the expressions for flavor violating processes. In other words, we relate the $d=6$ operators with $d=5$ by using a convenient parametrization \cite{Casas:2001sr, Ibarra:2003up}. The natural values of the Yukawa couplings for a low seesaw scale around the electroweak scale $\sim100$ GeV are of the order of $10^{-6}$. However, it turns out that there exists a portion of parameter space where the effect in these processes is observable, while neutrino masses remain small due to cancellations.

We investigate current bounds on these couplings from various processes in two minimal cases with two heavy neutrinos, a singlet and a triplet, and two triplets. It turns out that at least in these minimal cases, the number of parameters which specify the rate is reduced to a single real parameter, which is most constrained by the $\mu-e$ conversion experiments. We use this constraint to asses other possible channels and also comment on non-minimal models in the end.

We start with a discussion of the type I + III seesaw model of neutrino masses in section \ref{secMinModels}, where we focus on the two minimal cases and discuss the Casas-Ibarra-Ross parametrization. In section \ref{secMuEConversion}, we constrain the free parameters of the model using a bound from $\mu-e$ conversion searches in nuclei. Next, we consider a comprehensive list of other constraints in section \ref{secOther}, we comment on non-minimal models in section \ref{secBeyondMin} and present our conclusions and an outlook on future experiments in section \ref{secConclusions}. The appendices contain a derivation of the couplings of light and heavy leptons in models with arbitrary number of additional fermionic singlets and triplets and a calculation of $f_2 \to f_1 \gamma$ process in such models.

\section{Minimal type I and III models}
\label{secMinModels}

\begin{table}
\begin{center}
\begin{tabular}{lcc}
	\hline
	Parameter & Best fit & $3 \sigma$ \\ \hline
	$\Delta m_{21}^2[10^{-5} \text{eV}^2]$ 	& $7.65$ & $7.05-8.34$ \\
	$|\Delta m_{31}^2| [10^{-3} \text{eV}^2]$ 	& $2.40$ & $2.07-2.75$ \\
	$\sin^2 \theta_{12}$ 	& 0.304   & $0.25-0.37$ \\
	$\sin^2 \theta_{23}$ 	& 0.50   & $0.36-0.67$ \\
	$\sin^2 \theta_{13}$ 	& 0.01 & $\leq 0.056$ \\
	\hline
\end{tabular}
\end{center}
\caption{Parameter fits from oscillation experiments taken from \cite{Schwetz:2008er}.}
\label{TabNuData}
\end{table}

Neutrino oscillation experiments can be explained by non-zero masses of the light neutrinos. The best fit of the mass squared difference and the mixing angles are given in Table \ref{TabNuData} and constitute evidence for a nonzero mass. While the neutrino masses are bounded from above by beta decay searches and cosmology, the overall scale of the neutrino mass has not been established, therefore the lightest neutrino may still be massless. We will consider two minimal models which accommodate the oscillation data, one with a singlet and a triplet (motivated by a GUT) and the other with two triplets. The reason for this choice is minimality and predictivity. In both cases, the lightest neutrino is massless and there is only one Majorana phase which cannot be rotated away.

The Lagrangian for a model with a fermionic singlet and a zero hypercharge triplet can be written in the following way using the two component Weyl spinors
\begin{equation} \label{eqLagrIandIII}
\begin{split}
	\mathcal L_\ell &= 
	i L^\dagger_i \overline \sigma^\mu D_\mu L_i + i \ell_i^{c \dagger} \overline \sigma^\mu D_\mu \ell_i^c
	+ i T^\dagger_a \overline \sigma^\mu D_\mu T_a 
	+ i S^\dagger \overline \sigma^\mu \partial_\mu S \\
	& - \left( y_\ell^{ij} H^\dagger L_i  \ell_j^c - y_S^i H^T i \tau^2 L_i S - y_T^i H^T i \tau^2 \tau^a T_a L_i \right) +  \text{h.c.} \\
	& - 1/2 \left( m_T T^a T^a + m_S S S \right) + \text{h.c.}
\end{split}
\end{equation}
where $D_\mu$ stands for the appropriate covariant derivative. After spontaneous symmetry breaking, we obtain the well-known seesaw formula for the light neutrino masses
\begin{equation}
  (m^\nu)^{ij} = - \frac{v^2}{2} \left( \frac{y_T^i y_T^j}{m_T} + \frac{y_S^i y_S^j}{m_S} \right).
\end{equation}
When only type III is considered, the second term is replaced by the Yukawa couplings and the Majorana mass of the second triplet.

The same Yukawa couplings responsible for the $d=5$ operator, also contribute to the $d=6$ operators. For example, the presence of $y_S$ alters the couplings of the $W$ to the neutrino, which means that $U_{PMNS}$ is no longer unitary, while the $y_T$ mixes the charged leptons and therefore also affects the universality of $Z$ boson couplings. The Feynman rules for the fermion couplings in presence of a singlet and a triplet are presented in appendix \ref{appDerivCplngs}.

In order to use the information from the oscillation experiments on the neutrino mass to reduce the number of parameters, we employ a useful parametrization \cite{Ibarra:2003up}. For the two minimal cases, the neutrino masses are fixed, because the lightest neutrino is massless, therefore $\Delta m^2$ determines the mass of the heaviest two. This parametrization specifies all the Yukawa couplings in terms of measurable neutrino quantities and a single complex parameter $z$ for the case of normal (NH)
\begin{align} \label{eqYTNH}
  y_T^i &= - i \sqrt{2 m_T} / v \left(U_{i2} \sqrt{m_2^\nu} \cos z + U_{i3} \sqrt{m_3^\nu} \sin z \right)^*\,, 
  \\ \label{eqYSNH}
  y_S^i &= - i \sqrt{2 m_S} / v \left(-U_{i2} \sqrt{m_2^\nu} \sin z + U_{i3} \sqrt{m_3^\nu} \cos z \right)^*\,,
\end{align}
and inverted (IH) hierarchy
\begin{align} \label{eqYTIH}
  y_T^i &= - i \sqrt{2 m_T} / v \left(U_{i1} \sqrt{m_1^\nu} \cos z + U_{i2} \sqrt{m_2^\nu} \sin z \right)^*\,, 
  \\ \label{eqYSIH}
  y_S^i &= - i \sqrt{2 m_S} / v \left(-U_{i2} \sqrt{m_1^\nu} \sin z + U_{i2} \sqrt{m_2^\nu} \cos z \right)^*\,,
\end{align}
where $U$ is the unitary PMNS matrix defined by the standard parametrization and $\phi$ is the additional Majorana phase
\begin{equation}
\begin{split}
  U = 
  \begin{pmatrix}
    c_{12} c_{13} & s_{12} c_{13} & s_{13} e^{-i \delta} \\
    -s_{12} c_{23} - c_{12} s_{23} s_{13} e^{i \delta} & c_{12} c_{23} - s_{12} s_{23} s_{13} e^{i \delta}  & s_{23} c_{13} \\
    s_{12} s_{23} - c_{13} c_{23} s_{13} e^{i \delta} & - c_{12} s_{23} - s_{12} c_{23} s_{13} e^{i \delta} & c_{23} c_{13}
  \end{pmatrix} \\ \times \text{diag}{(1, e^{i \phi}, 1)}.
\end{split}  
\end{equation}
The size of the Yukawa couplings is determined by the complex $z$ parameter and it increases exponentially with $\text{Im}(z)$. In this case, the effects on the $d=6$ operators responsible for lepton flavor violating effects become visible, while at the same time neutrino masses remain small due to an exact cancellation. The higher the seesaw scale, the more severe fine-tuning is needed in order to produce a visible effect because the $d=6$ operators scale as $\Lambda^{-2}$ while neutrino masses go as $\Lambda^{-1}$.

The new particles may also be light, for example as predicted in \cite{Bajc:2006ia}. If this is so, we have the possibility to produce them at a high energy collider and measure the Yukawa couplings by decay rates and branching ratios \cite{Bajc:2007zf}. In this paper we instead investigate various lepton flavor violating (LFV) processes and determine the  values of $\text{Im}(z)$ which are needed in order to observe them.

When Yukawa couplings are large, $e^{\text{Im}(z)}$ factorizes, multiplies all the Yukawa couplings in Eqs.\eqref{eqYTIH}-\eqref{eqYSIH} and there is no dependency on the real component of $z$. Therefore, we can state all the limits on $d=6$ operators at a reference triplet mass $m_T$ with a single parameter, the imaginary part of $z$, which governs the overall size of $d=6$. The ratios between the various channels are not affected by $\text{Im}(z)$ and depend solely on neutrino mass parameters, together with the Majorana phase.
This means that, at least in the minimal models with two heavy neutrinos, the strictest bound in the $\mu e$ channel will put an upper limit on $\text{Im}(z)$ which suppresses also the other $\tau e$ and $\tau\mu$ channels.

\section{Constraints on Yukawa couplings from $\mu-e$ conversion in nuclei}
\label{secMuEConversion}

\begin{table}
  \begin{center}
  \begin{tabular}{lccc} \hline
    Nucleus                        & $V^{(p)} [m_\mu^{5/2}]$ & $V^{(n)} [m_\mu^{5/2}] $ & $\Gamma_{capture}[10^6 s^{-1}]$ \\ \hline
    $\text{Ti}^{48}_{22}$    &              0.0396            &             0.0468              &  2.59 \\
    $\text{Au}^{197}_{79}$ &              0.0974            &             0.146                &  13.07 \\ \hline
  \end{tabular}
  \end{center}
  \caption{Data taken from Tables I and VIII of \cite{Kitano:2002mt}.}
  \label{TabTiAuData}
\end{table}

The strictest bound on the $\mu e Z$ coupling is obtained by the $\mu-e$ conversion in a nucleus. The current bound on $Br_{\mu e} \equiv \Gamma_{conversion}/ \Gamma_{capture}$ was set by the SINDRUM collaboration from the experiments on titanium with $Br^{(Ti)}_{\mu e} < 4.3 \times 10^{-12}$ \cite{Dohmen:1993mp} and gold target setting the $Br_{\mu e}^{(Au)} < 7 \times 10^{-13}$ \cite{Bertl:2006up}, both at 90\%CL.

To get the constraint in the $\mu e$ channel from these experiments, one needs to know the expression for the rate in different nuclei. A detailed numerical calculation has been carried out by \cite{Kitano:2002mt} and we use their formula in Eq.(14) to calculate the desired conversion rate. The dominant contribution is due to a tree-level exchange of the $Z$ boson, the tree level Higgs amplitude being suppressed by the smallness of the charged lepton masses. Other contributions, involving also the singlet Yukawa couplings, are suppressed by a loop. Therefore at the leading order, the rate depends on the vectorial couplings only and using the notation of \cite{Kitano:2002mt} we have
\begin{equation}
  \Gamma_{conversion} = 2 G_F^2 \left[ 
  \left| \tilde g_{LV}^{(p)} V^{(p)} + \tilde g_{LV}^{(n)} V^{(n)} \right|^2 + 
  \left| \tilde g_{RV}^{(p)} V^{(p)} + \tilde g_{RV}^{(n)} V^{(n)} \right|^2
  \right]
\end{equation}
where $G_F$ is the SM Fermi coupling and $\tilde g_{L,RV}^{(p,n)}$ are found to be
\begin{align}
  \tilde g_{LV}^{(p)} &= 2 \left( 1- 4 s_w^2 \right) L_{12}^Z, & \tilde g_{RV}^{(p)} &= 2 \left( 1- 4 s_w^2 \right) R_{12}^Z, \\
  \tilde g_{LV}^{(n)} &= -2 L_{12}^Z, & \tilde g_{RV}^{(n)} &= -2 R_{12}^Z.
\end{align}
Throughout the paper we use $L^{W,Z}$ ($R^{W,Z}$) to denote the left (right) handed couplings of the fermions to the gauge bosons as defined in Eq.\eqref{eqLagAW} and $s_w = \sin \theta_w$, where $\theta_w$ is the weak mixing angle. The values of $V^{(p,n)}$ depend on the given nucleus and are specified in Table I of \cite{Kitano:2002mt}, while the capture rates are given in their Table VIII and we list the relevant quantities in Table \ref{TabTiAuData}. The resulting bound on the LFV couplings is
\begin{equation}
  |L_{12}^Z|^2 + |R_{12}^Z|^2 < 2.8\times 10^{-13},~2.3\times 10^{-14},
\end{equation}
for Ti and Au, respectively. Note that the bound due to the more recently measured Au channel is an order of magnitude stronger than the Ti bound previously considered in the literature \cite{Abada:2007ux, He:2009tf}. After allowing to vary the poorly known neutrino mass parameter $\theta_{13}$ within the allowed range in table 1 and the unknown phases $\delta$ and $\phi$, we obtain in the minimal models a bound on $\text{Im}(z) < 8.3 (7.9)$ for normal (inverted) hierarchy in case of one triplet and one singlet and $\text{Im}(z) < 8.0 (7.6)$ for two triplets, all at the reference mass of $m_T=100$ GeV for the lightest triplet.

When $\text{Im}(z)$ is so large, the branching ratios for decays of the triplets to light leptons are fixed by the neutrino mixing parameters and can be checked at the collider, if the triplet is light enough to be produced \cite{Arhrib:2009mz}. On the other hand, the same constraint puts an upper bound on the Yukawa coupling of the singlet, which makes it very hard to observe, even if it were light.

\section{Other constraints}
\label{secOther}

\begin{figure}[!ht]
\begin{center}
\includegraphics[width=10.5cm]{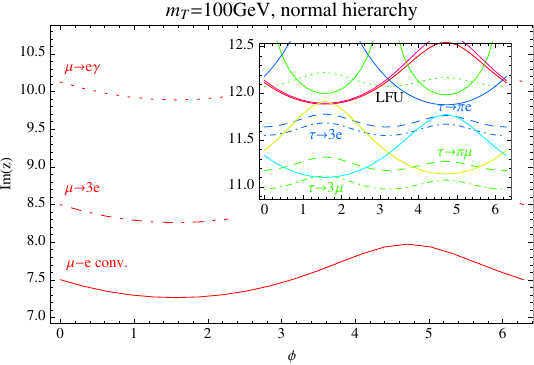}
\includegraphics[width=10.5cm]{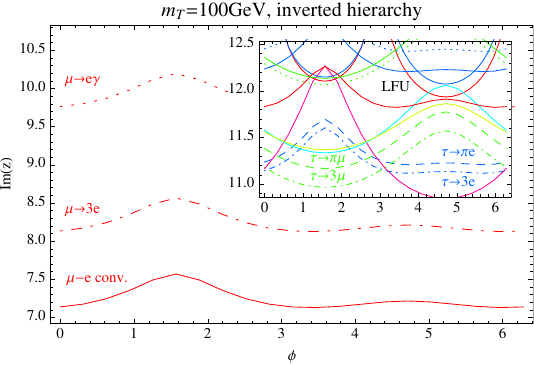}
\end{center}
\caption{ Comparison of various LFV and LFU bounds on the minimal type III model for normal (top) and inverted (bottom) neutrino mass hierarchy. The bounds coming from $\mu-e$ transitions are plotted in red, $\tau-e$ in blue and $\tau-\mu$ in green. Constraint from the $Z$ width to electrons is shown in magenta, to muons in cyan and to taus in yellow. The bounds constrain $\text{Im}(z)$ at the reference triplets' mass of $100$ GeV and depend on the unknown Majorana phase $\phi$. Dependence on the other poorly known neutrino parameters is negligible as explained in the text. \label{fig:1} }
\end{figure}

Previous phenomenological analyses of various experimental constraints on type III see-saw models \cite{Abada:2007ux, He:2009tf,Arhrib:2009mz} considered the three charged lepton flavor transitions separately. As explained above, such treatment is not necessarily justified, since the relative strengths of the various flavor transitions in type III are governed by neutrino mass and mixing parameters -- not all entries in the $L^{Z,W}_{ij}$, $R_{ij}^{Z,W}$ coupling matrices are independent. Consequently the impact of the various constraints should be compared through their bounds on the remaining free parameters of the model. In the case of the minimal III and I+III models this is the single complex parameter $z$. As explained in the next section, the general observation remains valid even in non-minimal type III models, albeit with more free parameters to be constrained. In the following we will compare the current and prospective bounds on the $\text{Im}(z)$ in the minimal III and I+III models. These are to be contrasted with the benchmark limits set by $\mu - e$ nuclear conversion experiments.  In most cases, the differences in bounds obtained with normal or inverted hierarchies and between III and I+III setups are not significant given their overall size. The bounds are not very sensitive to the angle $\theta_{13}$ and the Dirac phase $\delta$ due to the smallness of $\theta_{13}$. There is a mild sensitivity to the Majorana phase $\phi$ as shown on figure \ref{fig:1} where the comparison of various bounds is summarized for the minimal type III case.

Also to be kept in mind is that the dependence of observables on the size of $\text{Im}(z)$ is exponential, so that an improvement of a particular bound on $\text{Im}(z)$ by $\mathcal O(1)$ requires (numerically roughly two) orders of magnitude improvement in the actual experimental limit. On the other hand the LFV and lepton flavor universal (LFU) effects decouple quadratically with the lightest triplet mass as shown explicitly in appendix \ref{appDerivCplngs}.

Finally, the situation can also be viewed from the opposite perspective. Since within the I+III setup, the $\tau-\mu$ and $\tau-e$ LFV transitions for example, are constrained by the $\mu-e$ bound, any positive indication of the other transitions in the near future would indicate LFV contributions beyond the minimal models.

\underline{\it{Leptonic LFV decays}} are closely related to the $\mu-e$ conversion processes, since both receive dominant contributions form tree level Z exchange in type III models. The relevant decay widths at leading order are
\begin{subequations}
\begin{eqnarray}
  \Gamma_{\ell_i \rightarrow \ell_{j \neq k} \ell_{k} \overline \ell_k} &=& \frac{G_F^2 }{48 \pi^3} m_{\ell_i}^5
  \left( | L_{ij}^Z |^2 + | R_{ij}^Z |^2 \right) \left( | L_{kk}^Z |^2 + | R_{kk}^Z |^2 \right), 
  \\
  \Gamma_{\ell_i \rightarrow \ell_j \ell_{j} \overline \ell_j} &=& \frac{G_F^2 }{48 \pi^3} m_{\ell_i}^5 \left[ 
  \left( | L_{ij}^Z |^2 + | R_{ij}^Z |^2 \right) \left( | L_{jj}^Z |^2 + | R_{jj}^Z |^2 \right) \right. \nonumber
  \\
  && \hspace{1.8cm} \left.+ \frac{1}{2}\left(  | L_{ij}^Z  L_{jj}^Z |^2 + | R_{ij}^Z R_{jj}^Z |^2  \right)\right],
\end{eqnarray}
\end{subequations}
where we have neglected the final state lepton masses and doubly flavor suppressed amplitudes.

Taken the experimental limits from \cite{pdg}, it turns out that these decays constitute the most sensitive bounds on $\text{Im}(z)$ coming from $\tau-\ell$ transitions.

 \underline{\it{Radiative decays}} of the charged leptons also put limits on the LFV couplings. Since the photon coupling to the leptons remains universal at tree level, this process has to go through a loop. We have calculated the amplitudes coming from the $W$, $Z$ and Higgs loops and we give the result in the appendix \ref{secf2f1gamma}.

As seen in figure \ref{fig:1}, the limits coming from the loop suppressed $\mu \to e \gamma$ decay are substantially weaker than $\mu \to 3 e$, also due to a better experimental bound for the latter.

\underline{\it{Semileptonic LFV tau decays}}  $\tau\to\pi^0\ell$ and $\tau\to\eta\ell$, where $\ell=\mu,e$ were identified in \cite{He:2009tf,Arhrib:2009mz} as promising LFV signatures in the tau sector at low energies. Present experimental limits on the branching ratios are at the $10^{-8}$ level \cite{pdg} and thus the corresponding bounds could be in principle comparable to the ones from $\tau \to 3 \ell$. The decay widths induced by generic LFV $Z$ couplings can be written as
\begin{equation}
\begin{split}
  \Gamma_{\tau \rightarrow h \ell_i} &= \frac{ G_F^2 f_h^2}{8 \pi}
  m_{\tau}^3 \left( 1 - m_{h}^2/m_\tau^2 \right)^2 \left( |L_{3i}^Z|^2  + |R_{3i}^Z|^2  \right) \,,
\end{split}
\end{equation}
where $h=\pi^0,\eta$ and $f_h$ is the corresponding decay constant.

The above formula neglects final state lepton masses, but is accurate to a  percent level even for the $\tau\to\eta\mu$ channel. In our numerical analysis we use the complete kinematic formula, which can be found e.g. in \cite{He:2009tf} \footnote{We could reproduce all the $\pi$ and $\eta$ bounds in Table 1 of \cite{He:2009tf} except the one for $\tau\to\pi e$ for which we instead get in their notation $|\epsilon_{e\tau}|<6.0 \times 10^{-4}$. Therefore, this bound is not stronger than the one from $\tau \to 3 e$ decay as claimed in the paper.}. In our treatment of the hadronic matrix elements of the $\eta$ we follow the formalism of \cite{Feldmann:1998vh} and sum over contributions from all light quark flavours ($d\bar d$, $u\bar u$ and $s\bar s$) as also done in \cite{He:2009tf}. The derived limits on $\text{Im}(z)$ coming from $\pi^0$ and $\eta$ channels differ only slightly and exhibit identical $\phi$ dependence, therefore we do not plot them separately in figure \ref{fig:1}. 

\underline{\it{$Z$ decay widths}} to lepton pairs of various flavors were measured at LEP \cite{pdg}. Both flavor conserving (diagonal) as well as flavour changing (off-diagonal) decay modes could impose relevant constrains on deviations from the universal Z couplings to leptons. The relevant leading order decay width formula is 
\begin{equation}
\begin{split}
  \Gamma_ {Z \rightarrow \overline \ell_i \ell_j} &= \frac{G_F m_Z^3}{6 \pi \sqrt 2} \left( 
  \left( 1 - (m_{\ell_i} - m_{\ell_j})^2 / m_Z^2 \right) \left( 1 - (m_{\ell_i} + m_{\ell_j})^2 /m_Z^2 \right)  \right) \Large[ \\
  & \left( | L_{ij}^Z |^2 + | R_{ij}^Z |^2 \right) \left( 2 - (m_{\ell_i}^2 + m_{\ell_j} ^2 )/m_Z^2 - (m_{\ell_i}^2 - m_{\ell_j} ^2 )^2 /m_Z^4 \right) +  \\
  & 12 \text{ Re} \left( L_{ij}^Z R_{ij}^{Z*} \right) m_{\ell_i} m_{\ell_j} / m_Z^2 \Large ],
\end{split}
\end{equation}
where the finite lepton mass effects are only important for the tau channels.

Comparing to experimental measurements listed in \cite{pdg}, presently, the flavour diagonal channels yield bounds comparable to those from $\tau$ decays.

\underline{\it{Charged current lepton flavor universality tests}} can also probe for signs of violations of unitarity of the coupling matrix between light leptons, present in I+III models (the $3\times 3$ submatrix of $L^W_{ij}$). The most relevant observables here are (semi)leptonic kaon, pion and tau decays, while direct $W$ decay measurements at LEP yield somewhat weaker constraints. A model-independent analysis was performed in \cite{Abada:2007ux}. 
The best bounds on the deviations of unitarity are at the level of a few per-mille. When translated onto the bounds of the I+III model parameters, these are already quite weaker than other aforementioned constraints.

Additional constraints studied in the literature include anomalous magnetic moment of the muon  \cite{Biggio:2008in}, LFV leptonic and semileptonic decays of mesons, muonium -- anti-muonium oscillations  \cite{He:2009tf}, all of which yield much weaker constraints than the ones mentioned above.

\section{Beyond minimal models}
\label{secBeyondMin}

Before concluding, let us comment on non-minimal models with more than two heavy fermions. In the minimal cases above, we had 11 real parameters governing the Yukawa couplings: two Majorana masses of the heavy fermions, 5+2 parameters (the PMNS matrix and two masses of light neutrinos) mostly fixed from the oscillation data and finally, a single complex angle $z$ which specifies a complex orthogonal matrix $O$. Extending the model with another heavy fermion brings in another mass and another phase in the PMNS and also a third light neutrino mass (since the overall scale is unknown) and we now have 3 complex angles which specify the 3 by 3 orthogonal matrix $O$, altogether 18 parameters.

Although there are more free parameters in this case, correlations between different channels are generically preserved. This can easily be seen by considering the non-universal coupling,
\begin{equation}
\begin{split}
  L^Z_{e \mu} & \simeq \frac{v^2}{2} \sum_{\alpha=1}^{n_T}  y_{\alpha e}^* y_{\alpha \mu} / m_\alpha^2 \\
  &=  \sum_{\alpha=1}^{n_T} \sum_{i,j=1}^{3}\left( \sqrt{m_i^\nu m_j^\nu} / m_\alpha\right)  O_{\alpha i} O_{\alpha j} U_{ei} U_{\mu j} ,
\end{split}  
\end{equation}
where we sum over {\em all} the elements of the orthogonal matrix $O$, regardless of the flavor. Therefore one cannot easily enlarge the $\tau \ell Z$ couplings by enhancing a single element of $O$ without affecting the $\mu e$ channel and running in contradiction with the $\mu-e$ conversion experiment unless one aligns (fine-tunes) the available phases. This result holds for an arbitrary number of additional triplets and shows that the overall rate of the flavor processes is naturally dictated by the most constraining channel.

On the other hand, there is a potential gain in considering non-minimal models with three extra triplets. Namely, one can use the freedom of setting the overall scale of neutrinos at will and a positive signal is possible even for natural values of the Yukawas. For example, if light neutrinos are degenerate with the sum of their masses close to the upper limit from $\beta$ decay and cosmology  (say $\sum m_\nu \lesssim \text{eV}$ \cite{Strumia:2006db}), present $\mu - e$ conversion bounds already probe values of $\text{Im}(z_i) \simeq 3 - 5$.

\section{Conclusions and outlook}
\label{secConclusions}

\begin{figure}[t]
\begin{center}
\includegraphics[width=12cm]{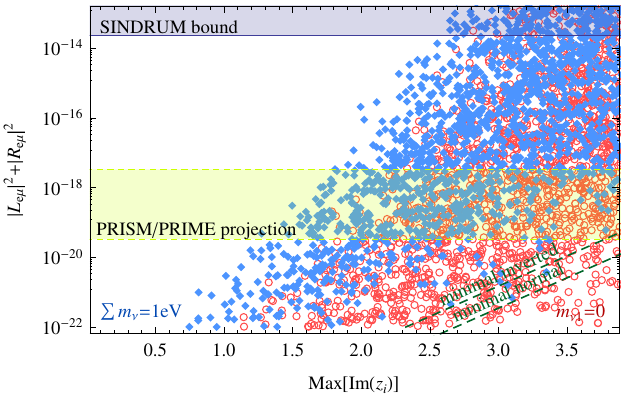}
\end{center}
\caption{Present and projected sensitivity of $\mu-e$ conversion experiments in non-minimal type III see-saw models for a massless lightest neutrino (in red empty circles, for both hierarchies) and for degenerate scenario at 1 eV (in blue filled spades). Minimal model predictions are drawn in green dashed lines. In all cases we put all the Majorana phases to zero and vary $z_i$ randomly. \label{fig:2} }
\end{figure}

The $\mu - e$ conversion limits will be further improved in the future by several orders of magnitude. According to proposals \cite{Ankenbrandt:2006zu} and \cite{P21Jparc, P20Jparc}, one can expect a sensitivity of $10^{-16}$ or even $10^{-18}$ by the PRISM/PRIME experiment. Such a sensitivity would constrain $\text{Im}(z)$ to 5.0 (4.6) in case of the minimal I + III model and to $4.6\, (4.2)$ for the minimal type III, again for normal (inverted) hierarchy.
For non-minimal models with degenerate eV scale neutrinos, these experiments would already probe $\text{Im}(z_i)\simeq 1 - 2$. Since the imaginary values of $z_i$ are free parameters of the model and setting any of them to zero does not enhance the symmetry of the Lagrangian, we consider such values natural. We plot both projections in figure \ref{fig:2} against the maximum value of $\text{Im}(z_i)$ in non-minimal models.

It is worth contrasting this with the projected sensitivity of the MEG experiment for the branching ratio of $\mu \to  e \gamma$, which is of the order $10^{-13}$ \cite{Dussoni:2009zz}. This corresponds to probing values of $\text{Im}(z)\sim 8.8$ in the minimal models.

Another interesting feature of a potential $\mu-e$ conversion signal would be its ability to distinguish the type III contribution from type I and II. This is due to the fact that the dependancy of the vectorial gauge boson couplings $V^{(p,n)}$ on the number of protons in the nucleus is different from the contributions of a scalar or a dipole operator (see \cite{Kitano:2002mt} for details). These may be present in type I and II, however they are loop suppressed and we do not consider them here. By measuring the conversion rate of two different nuclei, one can check for the vectorial nature which should dominate for the type III contribution.

The tau LFV decays are expected to be improved at a future flavor factory by one to two orders of magnitude \cite{Hitlin:2008gf}. The $Z$ leptonic width measurements could possibly be improved at the LHC or a future linear collider although we are not aware of existing dedicated sensitivity studies. On the other hand LFU tests in charged currents will be difficult to improve due to limiting theoretical uncertainties, although BESIII could improve on the present experimental precision \cite{Asner:2008nq}.

\section*{Acknowledgements}

M.N. would like to thank Borut Bajc and Goran Senjanovi\'c for discussions, encouragement, for careful reading of the manuscript and valuable suggestions. We also thank J\"orn Kersten for reading and helpful suggestions.  This work is supported in part by the Slovenian Research Agency, by the European Commission RTN network, Contract No. MRTN-CT-2006-035482 (FLAVIAnet) and by the Deutsche Forschungsgemeinschaft via the Junior Research Group ``SUSY Phenomenology" within the Collaborative Research Centre 676 ``Particles, Strings and the Early Universe''.

\appendix

\renewcommand{\thesection}{\Alph{section}}
\renewcommand{\theequation}{\Alph{section}.\arabic{equation}}
\section{Derivation of the fermion couplings} 
\label{appDerivCplngs}
\setcounter{equation}{0}

In this appendix we derive general expressions for the couplings of the charged and neutral fermions to the SM gauge and Higgs fields in the presence of a fermionic singlet and a weak triplet with hypercharge 0. We start with the Lagrangian in a two component notation and derive the rules in the four-component notation in the physical mass basis. In order not to clutter the notation, we initially consider an addition of a single fermionic triplet and a single singlet and we generalize the result for an arbitrary number of triplets $(n_T)$ and singlets $(n_S)$ in the end.

The starting point is the Lagrangian written with two component Weyl spinors in a basis where the Yukawa matrix of the charged fermions is real and diagonal and the Majorana masses $m_T$ and $m_S$ are also real
\begin{equation}\label{eqEffLagr}
\begin{split}
	\mathcal L_\ell &= 
	i L^\dagger_i \overline \sigma^\mu D_{L\mu} L_i
	 + i \ell^{c \dagger}_i \overline \sigma^\mu D_{\ell \mu} \ell^c_i
	+ i T^\dagger_a \overline \sigma^\mu D_{T \mu} T_a
	+ i S^\dagger \overline \sigma^\mu \partial_\mu S \\
	& - y_\ell^{ij} H^\dagger L_i  \ell_j^c + y_S^i H^T i \sigma^2 L_i S + y_T^i H^T i \sigma^2 \sigma^a T_a L_i +  \text{h.c.} \\
	& - 1/2 \left( m_T T_a T_a + m_S S S \right) + \text{h.c.}\,.
\end{split}
\end{equation}
Here, $i,j$ are the family indices running from 1 to 3 and the standard covariant derivatives are defined as
\begin{align}
  D_{L\mu}    &= \partial_\mu - i g/2 A_\mu^a \sigma^a - i g'/2 B_\mu, \\
  D_{\ell \mu} &= \partial_\mu + i g' B_\mu, \\
  D_{T \mu}   &= \partial_\mu + i g \varepsilon^{abc} T^b A_\mu^c,
\end{align}
where $a,b,c$ are the usual $SU(2)$ indices and $\sigma$'s are the Pauli matrices.

We use the linear combinations of the fields, labeled by their $U(1)$ charge $\sqrt 2 T^{\pm} = T^1 \mp i T^2$ and $T^0 = T^3$ and after spontaneous symmetry breaking, the Higgs field becomes
\begin{equation}
	H = 
	\begin{pmatrix}
	\phi^+ \\ (v + h + i \chi)/\sqrt{2}
	\end{pmatrix}\,,
\end{equation}
and the Lagrangian in Eq.\eqref{eqEffLagr} gives the following mass terms
\begin{equation} \label{eqMassTerms}
  \mathcal L_{mass} =  -
  \begin{pmatrix}
    \ell^c_i & T^+
  \end{pmatrix}
	M_\ell
  \begin{pmatrix}
    \ell_j
    \\
    T^-
  \end{pmatrix} -
  \begin{pmatrix}
    \nu_i & T_0 & S
  \end{pmatrix}
  M_\nu
  \begin{pmatrix}
    \nu_j \\
    T_0 \\
    S
  \end{pmatrix}/2 + \text{h.c.}\,,
\end{equation}
where
\begin{equation} \label{eqMassMatrices}
  M_\ell =
  \begin{pmatrix}
    v/\sqrt{2} \ y_{\ell}^{ij}\delta^{ij} & 0 \\
    v \ y_T^j & m_T
  \end{pmatrix} \text{ and }
  M_\nu = 
  \begin{pmatrix}
    0_{3 \times 3} & v \ y_T^i & v \ y_S^i \\
    v \ y_T^j & m_T & 0 \\
    v \ y_S^j & 0 & m_S
  \end{pmatrix}
\end{equation}
can be brought to a diagonal form by a biunitary and congruent transformation for the charged and neutral fields
\begin{equation}
  \label{eqBiunitary}
  \hat M_\ell = U^{+ \dagger} M_\ell U^-, \, \hat M_\nu = U^{0T} M_\nu U^0\,.
\end{equation}
In the limit when $v y_T \ll m_T$, one can expand these matrices in terms of small parameters $\varepsilon_i = v \, y_T^i / m_T$, $\varepsilon_{S i} = v \, y_S^i / m_S$,  and $\varepsilon'_i = v \, y_T^i \, m_i / m_T^2$,
\begin{align}
\label{eqExpUp}
  U^+ &=  \begin{pmatrix}
	1 - \frac{1}{2} |\varepsilon'_e|^2 &
	0 &
	0 &
	{\varepsilon'_e}^* \\
	0 &
	1 - \frac{1}{2} |\varepsilon'_\mu|^2 &
	0 &
	{\varepsilon'_\mu}^* 
	\\
	0 &
	0 &
	1 - \frac{1}{2} |\varepsilon'_\tau|^2 &
	{\varepsilon'_\tau}^* 
	\\
	- \varepsilon'_e &
	- \varepsilon'_\mu &
	- \varepsilon'_\tau &
	1 - \sum_i \frac{1}{2} \left| \varepsilon'_i \right|^2
  \end{pmatrix}, 
  \\
  \label{eqExpUm}
  U^- &=  
  \begin{pmatrix}
	1 - \frac{1}{2} |\varepsilon_e|^2 
	& - \frac{1}{2} \varepsilon_e^* \varepsilon_\mu
	& - \frac{1}{2} \varepsilon_e^* \varepsilon_\tau
	& \varepsilon_e^* 
	\\
	- \frac{1}{2} \varepsilon_e \varepsilon_\mu^*
	& 1 - \frac{1}{2} |\varepsilon_\mu|^2 
	& - \frac{1}{2} \varepsilon_\mu^* \varepsilon_\tau 
	& \varepsilon_\mu^*
	\\
	- \frac{1}{2} \varepsilon_e \varepsilon_\tau^*
	& - \frac{1}{2} \varepsilon_\mu \varepsilon_\tau^*
	& 1 - \frac{1}{2} |\varepsilon_\tau|^2 
	& \varepsilon_\tau^*
	\\
	- \varepsilon_e 
	& - \varepsilon_\mu
	& - \varepsilon_\tau 
	& 1 - \frac{1}{2} \sum_i |\varepsilon_i|^2
  \end{pmatrix},
\\
  U^0 &= 
  \begin{pmatrix}
   (\delta_{ik} - \frac{1}{4}( \varepsilon_i^* \varepsilon_k + \varepsilon_{S i}^* \varepsilon_{S k} )) U_{kj} & \varepsilon_j^* / \sqrt 2 & \varepsilon_{S j}^*/\sqrt2 
   \\
    - \varepsilon_k U_{kj} / \sqrt 2		 & 1 - \frac{1}{4} \sum_i | \varepsilon_i |^2 & \sum_i \varepsilon_i \varepsilon_{S i}^* 
    \\
    - \varepsilon_{S k} U_{kj} / \sqrt 2	& \sum_i \varepsilon_i^* \varepsilon_{S i}	& 1 - \frac{1}{4} \sum_i |\varepsilon_{S i}|^2 
  \end{pmatrix}.
\end{align}

After performing these rotations, we combine the mass eigenstates of the charged fermions and the triplets into a four component Dirac spinor while the neutral fermions form a Majorana spinor using the usual prescription
\begin{equation}
  \ell_i = \begin{pmatrix} \ell_i \\ \overline{\ell^c}_i \end{pmatrix}, \
  T^- = \begin{pmatrix} T^- \\  \overline{T^+} \end{pmatrix}, \
  \nu_i = \begin{pmatrix} \nu_i \\ \overline{\nu}_i \end{pmatrix}, \
  T^0 = \begin{pmatrix} T^0 \\ \overline{T^0} \end{pmatrix}.
\end{equation}
The mixing matrices alter the gauge couplings of the SM fermions and since they mix the chiral fermions with vector-like triplets, it is convenient to introduce a general notation for the charged and neutral four component spinors (see also the appendix of \cite{Arhrib:2009mz})
\begin{align}
  f^-_i = (e, \mu, \tau, T^-), \   f^0_j = (\nu_1, \nu_2, \nu_3, T^0, S).
\end{align}
Using such a convention, we can write down the $W$ and $Z$ couplings in a unified way with
\begin{equation}
	\gamma^\mu = 
	\begin{pmatrix}
		0 & \sigma^\mu \\
		\overline \sigma^\mu & 0\\
	\end{pmatrix},
	\,
	\sigma^\mu = (1, -\sigma^i),
	\,
	\gamma^5 = 
	\begin{pmatrix}
	1 & 0 \\
	 0 & -1
	\end{pmatrix},
	\,
	P_{L,R} = \frac{1 \pm \gamma^5}{2}\,,
\end{equation}
and we have a Lagrangian
\begin{align} 
    \label{eqLagAW}
    \mathcal L_{int} &= 
    - e \, \overline f_i \slashed A  f_i +
    \left(g \, \overline f'_i \slashed{W}^+ (L^W P_L + R^W P_R)_{ij} f_j + \text{h.c.} \right) +
    \\ \label{eqLagPhi}
    &
    \left( \phi^+ \overline f'_j \left(L^{\phi} P_L +  R^{\phi} P_R \right)_{ji} f_i + \text{h.c.} \right) +
    \\ \label{eqLagZ}
    & \frac{g}{c_w} \, \overline f_i \slashed Z (L^Z P_L + R^Z P_R)_{ij} f_j +
    \chi \overline f_i \left(L^\chi P_L + R^\chi P_R \right)_{ij} f_j
    \\
    & 
    + h \overline f_i \left(L^h P_L + R^h P_R \right)_{ij} f_j\,,
\end{align}
with the following gauge
\begin{align}
  \label{eqLRw}
  L^W_{ij} &= U^{0*}_{\alpha i} U^-_{\alpha j} / \sqrt 2 + U^{0*}_{\beta i} U^-_{\beta j},
  & 
  R^W_{ij} &= U^0_{\beta i} U^+_{\beta j},
  \\ \label{eqLRz}
  L^Z_{ij}  &= (s_w^2 - 1/2) U^{- *}_{\alpha i} U^-_{\alpha j} - c_w^2 U^{-*}_{\beta i} U^-_{\beta j},
  & 
  R^Z_{ij} &= s_w^2 U^{+ *}_{\alpha i} U^+_{\alpha j} - c_w^2 U^{+ *}_{\beta i} U^+_{\beta j},
\end{align}
and would-be-Goldstone and physical Higgs couplings
\begin{align}
	L^\phi_{ij} &= y_T^{\beta-3 \alpha} \left( \sqrt 2 U^0_{\alpha i} U^-_{\beta j} -
	U^0_{\beta i} U^-_{\alpha j} \right) + y_S^{\gamma-3-n_T \alpha} U^0_{\gamma i} U^-_{\alpha j},
	\\
	R^\phi_{ij} &= - y_\ell^{\alpha} U^{0*}_{\alpha i} U^+_{\alpha j},
	\\
	L^\chi_{ij} &= \frac{i}{\sqrt 2} y_{\ell}^{\alpha} U^{+*}_{\alpha i} U^-_{\alpha j} - i y_T^{\beta-3 \alpha} U^{+*}_{\beta i} U^-_{\alpha j},
	\\
	R^\chi_{ij} &= - \frac{i}{\sqrt 2} y_\ell^{\alpha} U^{-*}_{\alpha i} U^+_{\alpha j} + i y_T^{\beta-3 \alpha *} U^{-*}_{\alpha i} U^+_{\beta j}.
	\\
	L^h_{ij} &= - \frac{1}{\sqrt 2} y_\ell^{\alpha} U^{+*}_{\alpha i} U^-_{\alpha j} - y_T^{\beta-3 \alpha} U^{+*}_{\beta i} U^-_{\alpha j},
	\\
	R^h_{ij} &= - \frac{1}{\sqrt 2} y_\ell^{\alpha} U^{-*}_{\alpha i} U^+_{\alpha j} - y_T^{\beta-3 \alpha*} U^{-*}_{\alpha i} U^+_{\beta j}.
\end{align}
In the notation above, repeated indices are always summed over. The indices $\alpha, \alpha'$ run over the light families from 1 to 3, $\beta$ runs over the number of triplets from 4 to $3 + n_T$, while $\gamma$ is the singlet index going from $4 + n_T$ to $3 + n_T + n_S$. When additional copies of particles are considered, the mass matrices in Eq.\eqref{eqMassMatrices} have to be extended.

A couple of features of the model are noteworthy. While the photon vertex remains universal at the tree-level, the $Z$ vertex now receives off-diagonal entries. Also, the right-handed couplings are now present, however they are always suppressed by the mass of the light charged fermions $m_\ell/m_T$ which can be seen from the expansion of $U^+$ in Eq.\eqref{eqExpUp}. Notice that the SM limits are easily obtained, by either sending $y_{T,S} \to 0$ and/or $m_{T,S} \to \infty$. In this case, the mixing matrices become diagonal and the SM expressions are recovered.

\section{$f_2 \to f_1 \gamma$ calculation}
\label{secf2f1gamma}
\setcounter{equation}{0}

Here, we discuss the calculation of amplitudes for the $f_2 \to f_1 \gamma$ decay. We have done the calculation in $R_\xi$ gauge with arbitrary left and right-handed gauge couplings of the fermions with arbitrary masses $m_{1,2}$ of $f_{1,2}$. The amplitude is proportional to the $d=5$ operator
\begin{equation} \label{EqMagTrans}
	i \sigma_{\mu \nu} \varepsilon^\mu q^\nu, 
\end{equation}
where $\varepsilon$ is the polarization vector of the photon with momentum $q = p_2 - p_1$ and $p_i$ are the four-momenta of $f_i$. 
The final result has to be finite and $\xi_{w,z}$ independent.

Before giving the transition amplitude, we would like to comment on the divergency cancellations in models with non-unitary mixing matrices which is the case for the type III seesaw. If the $W$ coupling matrix is unitary, the divergent part proportional to $\slashed \varepsilon$ vanishes or it is cancelled by the diagrams with photons radiating from the external fermions. This does not happen in models where vector-like fermions mix with the chiral. The problem is resolved by noting that
\begin{equation} \label{eqNonUnitWNonUnivZ}
	\left( L^{W\dagger} L^W \right)_{ij} = -L^Z_{ij}, \quad i \neq j,
\end{equation}
which is a consequence of the $SU(2)$ structure of the electroweak Lagrangian. The relation in Eq.\eqref{eqNonUnitWNonUnivZ} holds also for $R$ couplings, and both can be checked from Eqs.\eqref{eqLRw} and \eqref{eqLRz}. Given that the non-unitarity of the mixing matrix is directly related to the non-universality of the $Z$ coupling, we expect the cancellation to come from a diagram with a single off-diagonal $Z$ coupling. Indeed, when we calculate the $Z-\gamma$ mixing diagrams, the divergent part vanishes.

In order to get a finite and gauge invariant result, we have to sum the diagrams with unphysical would-be-Goldstone fields. Their couplings can be related to the gauge boson couplings (see also \cite{Lavoura:2003xp})
\begin{align}
   L^{\phi}_{ij} &= g / m_W \left( L^W_{ij} m_i - R^W_{ij} m_j \right), \\
   L^{\chi}_{ij} &= g / i c_w m_Z \left( L^Z_{ij} m_i - R^Z_{ij} m_j \right),
\end{align}
and symmetrically for $R \to L$. Using these relations, all the $\xi$ dependent terms cancel to all orders in $m_{1,2}$ and we have a finite, gauge invariant result coming from the $W$, $Z$ and Higgs loops, together with corresponding $\phi$ and $\chi$ loops. We expand the scalar integrals in small $m_2$, set $m_1 = 0$ and get the amplitudes

\begin{align}
	\nonumber \label{eqAmpWR}
	\mathcal M^W_R &= \frac{G_F}{\sqrt 2} \frac{e}{24 \pi^2} \overline f_1 i \sigma_{\mu \nu} \varepsilon^\mu q^\nu P_R f_2
	 \sum_{n=1}^{3 + n_T + n_S} \frac{1}{(1 - x_n)^{4}}\Bigl [
	\\
	&6 m_n L^{W*}_{n1} R^W_{n2} \left( x_n^3 - 12 x_n^2 + 6 x_n^2 \log x_n + 15 x_n - 4 \right) \left( 1 - x_n \right) +
	\\ \nonumber
	&m_2 L^{W*}_{n1} L^W_{n2} \left( 4 x_n^4 + 18 x_n^3 \log x_n - 49 x_n^3 + 78 x_n^2 - 43 x_n + 10 \right) \Bigr],
	\\ \nonumber
	\mathcal M^Z_R &= - \frac{G_F}{\sqrt 2} \frac{e}{24 \pi^2} \overline f_1 i \sigma_{\mu \nu} \varepsilon^\mu q^\nu P_R f_2
	\sum_{c=1}^{3+n_T} \frac{1}{(1 - x_c)^{4}} \Bigl [ 
	\\
	&6 m_c L^Z_{1c} R^Z_{c2} \left( x_c^3 - 6 x_c \log x_c + 3 x_c - 4 \right) \left(1 - x_c \right)+
	\\ \nonumber
	&m_2 L^Z_{1c} L^Z_{c2} \left( 5 x_c^4 - 14 x_c^3 - 18 x_c^2 \log x_c  + 39 x_c^2 - 38 x_c + 8 \right) \Bigr],
	\\ \nonumber
	\mathcal M^h_R &= -\frac{e}{96 \pi^2 m_h^2} \overline f_1 i \sigma_{\mu \nu} \varepsilon^\mu q^\nu P_R f_2
	\sum_{c=1}^{3+n_T} \frac{1}{(1 - y_c)^{4}} \Bigl [
	\\
	& 6 m_c R^h_{1c} R^h_{c2} \left( y_c^2 - 4 y_c + 2 \log y_c + 3 \right) \left(1 - y_c \right) -
	\\ \nonumber
	& m_2 R^h_{1c} L^h_{c2} \left( y_c^3 - 6 y_c^2 + 6 y_c \log y_c + 3 y_c + 2) \right)
	\Bigr],
\end{align}

\noindent
where $n$ sums over the neutral particles in the loop (three light neutrinos and $n_T$ + $n_S$ heavy mediators), $c$ sums over the charged particles (three light $e, \mu, \tau$ and $n_T$ heavy triplets) and $x_n = m_n^2 / m_W^2$, $x_c = m_c^2 / m_Z^2$, $y_c = m_c^2 / m_h^2$. The amplitude proportional to $P_L$ is obtained by substituting $(L,R) \to (R,L)$. Finally, the total decay rate for the process is given by
\begin{equation}
	\Gamma_{f_2 \to f_1 \gamma} = \frac{m_2^3}{16 \pi} \left( |\mathcal M_L|^2 + |\mathcal M_R|^2  \right),
\end{equation}
with $\mathcal M_{L,R} = \sum_{i=W,Z,h} \mathcal M_{L,R}^i$.

Notice that the result above is valid also for theories with right-handed gauge couplings, e.g. in left-right symmetric theories. From the results above, one can easily reproduce the calculations for type I case. The $Z$ and $H$ amplitudes are zero, so are the $R^W$ couplings, therefore the only contributing piece is the third line of Eq.\eqref{eqAmpWR}, proportional to $L^W_{ij} = U^{0*}_{ij} / \sqrt2$. With this substitution we reproduce the well-known results in \cite{Cheng:2000ct}. We cannot fully reproduce the results of \cite{Abada:2008ea} for the case of pure type III, our result for the decay rate is bigger by roughly a factor of two.


\end{document}